\documentclass[fleqn,usenatbib]{mnras}

\usepackage{newtxtext,newtxmath}

\usepackage[T1]{fontenc}

\DeclareRobustCommand{\VAN}[3]{#2}
\let\VANthebibliography\thebibliography
\def\thebibliography{\DeclareRobustCommand{\VAN}[3]{##3}\VANthebibliography}

\usepackage{graphicx}	
\usepackage{amsmath}	
\usepackage[flushleft]{threeparttable}

\def\hst{\textit{HST}}
\def\planck{\textit{Planck}}

\def\kms {\rm km\,s^{-1}}
\def\kmsmpc{\rm km\,s^{-1}\,Mpc^{-1}}

\newcommand\rxj{RXJ\,1131$-$1231}
\newcommand\he{HE\,0435$-$1223}
\newcommand\pg{PG\,1115$+$080}

\newcommand\jj{J\,0924+0219}
\newcommand{\sref}[1]{Section~\ref{#1}}
\newcommand{\fref}[1]{Figure~\ref{#1}}
\newcommand{\tref}[1]{Table~\ref{#1}}
\newcommand{\eref}[1]{Equation~(\ref{#1})}

\newcommand{\DsDds}{D_{\textrm{s}}/D_{\textrm{ds}}}
\newcommand{\Dd}{D_{\textrm{d}}}
\newcommand{\Ds}{D_{\textrm{s}}}
\newcommand{\Dds}{D_{\textrm{ds}}}
\newcommand{\Ddt}{D_{\Delta\textrm{t}}}
\newcommand{\lambdaint}{\lambda_{\textrm{int}}}
\newcommand{\kappaext}{\kappa_{\textrm{ext}}}

\title[J0924 mass distribution and time delay prediction]{SHARP VIII: \jj\ lens mass distribution and time-delay prediction through adaptive-optics imaging}

\author[Geoff~C.-F.~Chen et al.]{
Geoff~C.-F.~Chen,$^{1}$\thanks{Current E-mail: gcfchen@astro.ucla.edu}
Christopher~D.~Fassnacht,$^{2}$
Sherry~H.~Suyu,$^{3,4,5}$
\newauthor{
L{\'e}on~V.~E.~Koopmans,$^{6}$
David~J.~Lagattuta,$^{7,8}$
John~P.~McKean,$^{6,9}$
Matt~W.~Auger,$^{10}$
}
\newauthor{
Simona~Vegetti$^{3}$
and Tommaso Treu$^{1}$
}
\\
$^{1}$Department of Physics and Astronomy, University of California, Los Angeles, CA 90095, USA\\
$^{2}$Department of Physics and Astronomy, University of California, Davis, CA 95616, USA\\
$^{3}$Max $\planck$ Institute for Astrophysics, Karl-Schwarzschild-Strasse 1, D-85740 Garching, Germany\\
$^{4}$Technische Universität München, Physik-Department, James-Franck-Str. 1, 85748 Garching, Germany\\
$^{5}$Academia Sinica Institute of Astronomy and Astrophysics (ASIAA), 11F of ASMAB, No.1, Section 4, Roosevelt Road, Taipei 10617, Taiwan \\
$^{6}$Kapteyn Astronomical Institute, University of Groningen, P.O.Box 800, 9700 AV Groningen, The Netherlands\\
$^{7}$Centre for Extragalactic Astronomy, Durham University, South Road, Durham DH1 3LE, UK\\
$^{8}$Institute for Computational Cosmology, Durham University, South Road, Durham DH1 3LE, UK\\
$^{9}$ASTRON, Netherlands Institute for Radio Astronomy, P.O. Box 2, 7990 AA Dwingeloo, the Netherlands\\
$^{10}$Institute of Astronomy, University of Cambridge, Madingley Rd, Cambridge, CB3 0HA, UK\\
}

\date{Accepted XXX. Received YYY; in original form ZZZ}

\pubyear{2015}

\begin{document}
\label{firstpage}
\pagerange{\pageref{firstpage}--\pageref{lastpage}}
\maketitle

\begin{abstract}
Strongly lensed quasars can provide measurements of the Hubble constant ($H_{0}$) independent of any other methods. 
One of the key ingredients is exquisite high-resolution imaging data, such as Hubble Space Telescope (HST) imaging and adaptive-optics (AO) imaging from ground-based telescopes, which provide strong constraints on the mass distribution of the lensing galaxy. 
In this work, we expand on the previous analysis of three time-delay lenses with AO imaging (\rxj, \he, and \pg), and perform a joint analysis of \jj~by using AO imaging from the Keck Telescope, obtained as part of the SHARP (Strong lensing at High Angular Resolution Program) AO effort, with HST imaging to constrain the mass distribution of the lensing galaxy. 
Under the assumption of a flat $\Lambda$CDM model with fixed $\Omega_{\rm m}=0.3$, we show that by marginalizing over two different kinds of mass models (power-law and composite models) and their transformed mass profiles via a mass-sheet transformation,
we obtain
$\Delta t_{\rm BA}h\hat{\sigma}_{v}^{-2}=6.89\substack{+0.8\\-0.7}$ days, $\Delta t_{\rm CA}h\hat{\sigma}_{v}^{-2}=10.7\substack{+1.6\\-1.2}$ days, and $\Delta t_{\rm DA}h\hat{\sigma}_{v}^{-2}=7.70\substack{+1.0\\-0.9}$ days, where $h=H_{0}/100~\kmsmpc$ is the dimensionless Hubble constant and $\hat{\sigma}_{v}=\sigma^{\rm ob}_{v}/(280~\kms)$ is the scaled dimensionless velocity dispersion. Future measurements of time delays with 10\% uncertainty and velocity dispersion with 5\% uncertainty would yield a $H_0$ constraint of $\sim15$\% precision.
\end{abstract}




\section{Introduction}
Measuring the Hubble constant is one of the most important tasks in modern cosmology especially since not only it sets the age, the size, and the critical density of the Universe but also the recent direct $H_{0}$ measurements from Type Ia supernovae (SNe), calibrated by the traditional Cepheid distance ladder \cite[SH0ES collaboration;][]{RiessEtal19}, show a $4.4\sigma$ tension with the Planck results under the assumption of $\Lambda$CDM model \citep[e.g.,][]{KomatsuEtal11,HinshawEtal13,planck18parameter,AndersonEtal14,KazinEtal14,RossEtal15}.  
However, a recent measurement of $H_{0}$ from SNe Ia calibrated by the Tip of the Red Giant Branch (TRGB) by the Carnegie-Chicago Hubble Program (CCHP) agrees with both the Planck and SH0ES results within the errors \citep{FreedmanEtal19,FreedmanEtal20}.
These results clearly demonstrate that it is crucial to test any single measurement by independent probes.

Strongly lensed quasars provide an independent way to measure the Hubble constant \citep{Refsdal64,SuyuEtal17,TreuMarshall16}. 
With the combination of time delays, high-resolution imaging, the velocity dispersion of the lensing galaxy, and the description of the mass along the line of sight \citep[so-called external mass-sheet transformation; see details in][]{FalcoEtal85,GorensteinEtal88,FassnachtEtal02,SuyuEtal13,GreeneEtal13,CollettEtal13}, the TDCOSMO\footnote{\url{http://www.tdcosmo.org/}} collaboration \citep[][]{MillonEtal20} has shown that one can provide robust constraints on both the angular diameter distance to the lens \citep[$\Dd$;][]{JeeEtal15} and the time-delay distance which is a ratio of the angular diameter distances in the system: 
\begin{equation}
\label{eq:theory7}
\Ddt\equiv\left(1+
z_{\rm d}\right)\frac{\Dd\Ds}{\Dds}\propto H_{0}^{-1},
\end{equation}
where $z_{\rm d}$ is the redshift of the lens, $\Ds$ is the distance to the background source, and $\Dds$ is the distance between the lens and the source. These distances are used to determine cosmological parameters, primarily $H_0$ \citep[e.g.,][]{SuyuEtal14,BonvinEtal16,BirrerEtal19,GChenEtal19,RusuEtal19, WongEtal20,JeeEtal19,TaubenbergerEtal19,ShajibEtal20_0408}.

A blind analysis done by \citet{WongEtal20} with this technique as part of the H0LiCOW program \citep[][]{SuyuEtal17}, in collaboration with the COSMOGRAIL \citep[e.g.,][]{CourbinEtal18} and SHARP \citep[][Fassnacht et al. in prep]{GChenEtal19} programs, combined the data from six gravitational lens systems\footnote{Except the first lens, B1608+656, which was not done blindly, the subsequent five lenses in H0LiCOW are analyzed blindly with respect to the cosmological quantities of interest.}, and inferred $H_0 = 73.3\substack{+1.7\\-1.8}~\kmsmpc$, a value that was $3.8\sigma$ away from the Planck results. 
The above work marginalized over two different kinds of mass profiles for the lensing galaxies in order to better estimate the uncertainties. 
The first description consists of a NFW dark matter halo \citep{NavarroEtal96} plus a constant mass-to-light ratio stellar distribution (the ``composite model'').  
The second description models the three dimensional total mass density distribution, i.e., luminous plus dark matter, of the galaxy as a power law \citep[][]{Barkana98}, i.e., $\rho(r) \propto r^{-\gamma}$ (the power-law model). 
\citet{MillonEtal20} later combined six lenses from \citet{WongEtal20} with one additional lens analyzed by \citet{ShajibEtal20_0408} in the STRIDES program \citep[][]{TreuEtal18}, and showed that even if we separate these two descriptions of the mass distribution of the lensing galaxy, the $H_0$ measurements are consistent well within 1\%. An independent check by \citet{GChenEtal19} using ground-based high-resolution adaptive optics (AO) imaging data from SHARP\footnote{The Keck AO imaging data are part of the Strong-lensing High Angular Resolution Programme (SHARP; Fassnacht et al. in preparation)} with three strongly lensed quasar also shows consistent results with \citet{WongEtal20} and is 3.5$\sigma$ away from Planck results. 

Given the growing statistical tension between $H_{0}$ measurements, efforts by the TDCOSMO collaboration have gone into studying potential systematic uncertainties \citep{MillonEtal20,GilmanEtal20}. A crucial potential source of uncertainty is the assumptions on the radial density profile. 
\citet{BirrerEtal20} introduced a flexible parametrization on the mass model that is maximally degenerate with $H_0$
%
%
through the mass-sheet trasformation \citep[so-called internal MST; see also][]{SchneiderSluse13,XuEtal16,Kochanek20,Kochanek21,GChenEtal20}, as a way to express departures from the standard assumptions in previous work \citep{BlumEtal20, ShajibEtal21}. 
With this parametrization, the main factor determining the precision of the cosmological inference is the stellar kinematics in the lensing galaxy \citep[see discussion by][]{TreuKoopmans02,KoopmansEtal03b,JeeEtal16,BirrerEtal16,GChenEtal20}.
%
%
With the MST parametrization, the uncertainty on $H_0$ based on the 7 lens sample of \citet{MillonEtal20} goes from $\sim$2\% to $\sim8\%$, in a standard $\Lambda$CDM cosmology.

To further constrain the $H_{0}$ value contributed from the MST and anisotropy parameters, 
\citet{BirrerEtal20} developed a hierarchical Bayesian framework 
by including external datasets, assuming they are drawn from the same population.
When assuming that the TDCOSMO lenses and the SLACS samples are drawn from the single stellar-orbit anisotropy distribution \citep[][]{BoltonEtal04SDSS,BoltonEtal06SDSS,AugerEtal10}, \citet{BirrerEtal20} inferred $73.3\pm5.8~\kmsmpc$. 
Assuming that TDCOSMO and SLACS are also drawn from the same population in terms of both anisotropy and mass density profile, the inference on $H_{0}$ shifted to 
$67.4\substack{+4.1\\-3.2}~\kmsmpc$, which statistically agree with both Planck and SH0ES results. 
Increasing the number of the time-delay lens systems and using different external datasets are crucial to assess whether the difference between SLACS and TDCOSMO is real or a statistical fluctuation \citep{BirrerTreu21}.


To expand the sample of analyzed AO-observed time-delay lenses \citep{GChenEtal19} 
we study the \jj~lens system, which has AO imaging and archival HST imaging. 
In this work, we take into account both the internal and external MST and forecast the time delays. 
Since the velocity dispersion of the lensing galaxy is not yet measured, we predict the time delay based on the imaging data with the expected precision of the kinematic data. 
In \sref{sec:data}, we describe the basic information on \jj~and describe the data acquisition and analysis; in \sref{sec:jjmodeling}, we describe the models we used for fitting the imaging; 
in \sref{sec:predictedTD_LCDM}, we make a time-delay prediction based on imaging data under the assumption of a flat $\Lambda$CDM model with fixed $\Omega_{\rm m}=0.3$. The conclusion is in \sref{sec:conclusion}.

\section{\jj}
\label{sec:data}

\begin{figure*}
    \includegraphics[width=\linewidth]{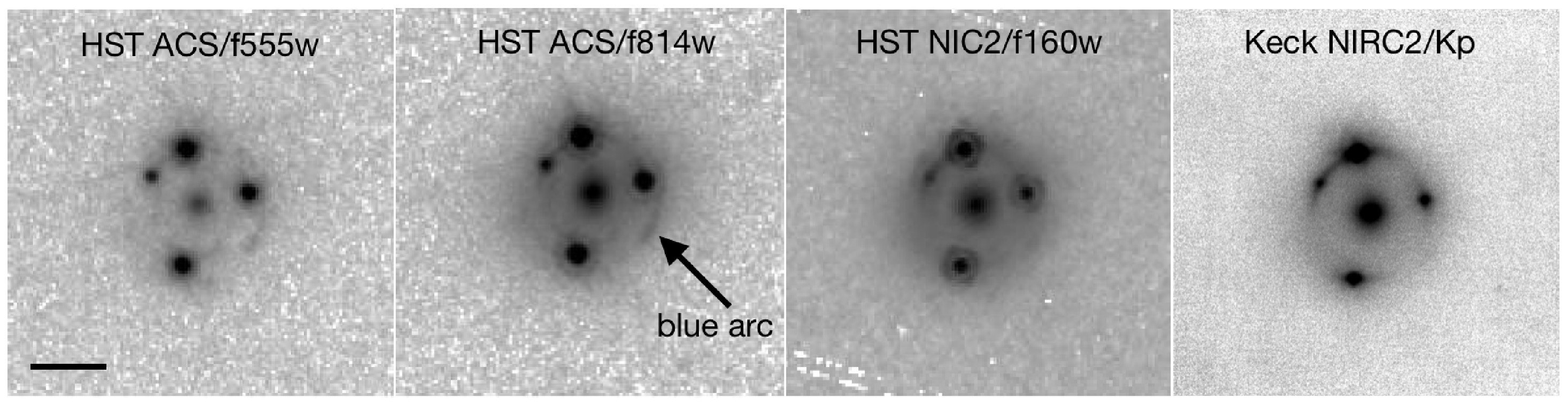}
    \caption{HST and Adaptive optics images of \jj~gravitational lens systems. The solid horizontal line represents 1\arcsec scale. The foreground main lens is located in the center of the lens system. The multiple lensed images and the extended arc around the lensing galaxy are from the background AGN and its host galaxy.}
\label{fig:aoimages}
\end{figure*}

The \jj\ system (J2000: 09
$^{\textrm{h}}$24$^{\textrm{m}}$55.87, 
02\degr19\arcmin24\farcs9) is a quadruply-lensed quasar discovered by 
\citet{InadaEtal03}. The main lensing galaxy is at a redshift of $z_{\rm d} = 0.394\pm0.001$ 
\citep{EigenbrodEtal06}, and the source redshift is $z_{\rm s} = 1.524\pm0.001$ 
\citep{InadaEtal03}.
The analysis in this paper is based on new Keck AO and archival \hst\ observations of \jj. We describe the data acquisition and analysis in \sref{sec:HSTreduction} and \sref{sec:AOreduction}.
We show the data from three HST bands and one Keck AO K$^\prime$-band in Figure~\ref{fig:aoimages}.
    
\begin{figure*}
    \centering
    \includegraphics[width=\linewidth]{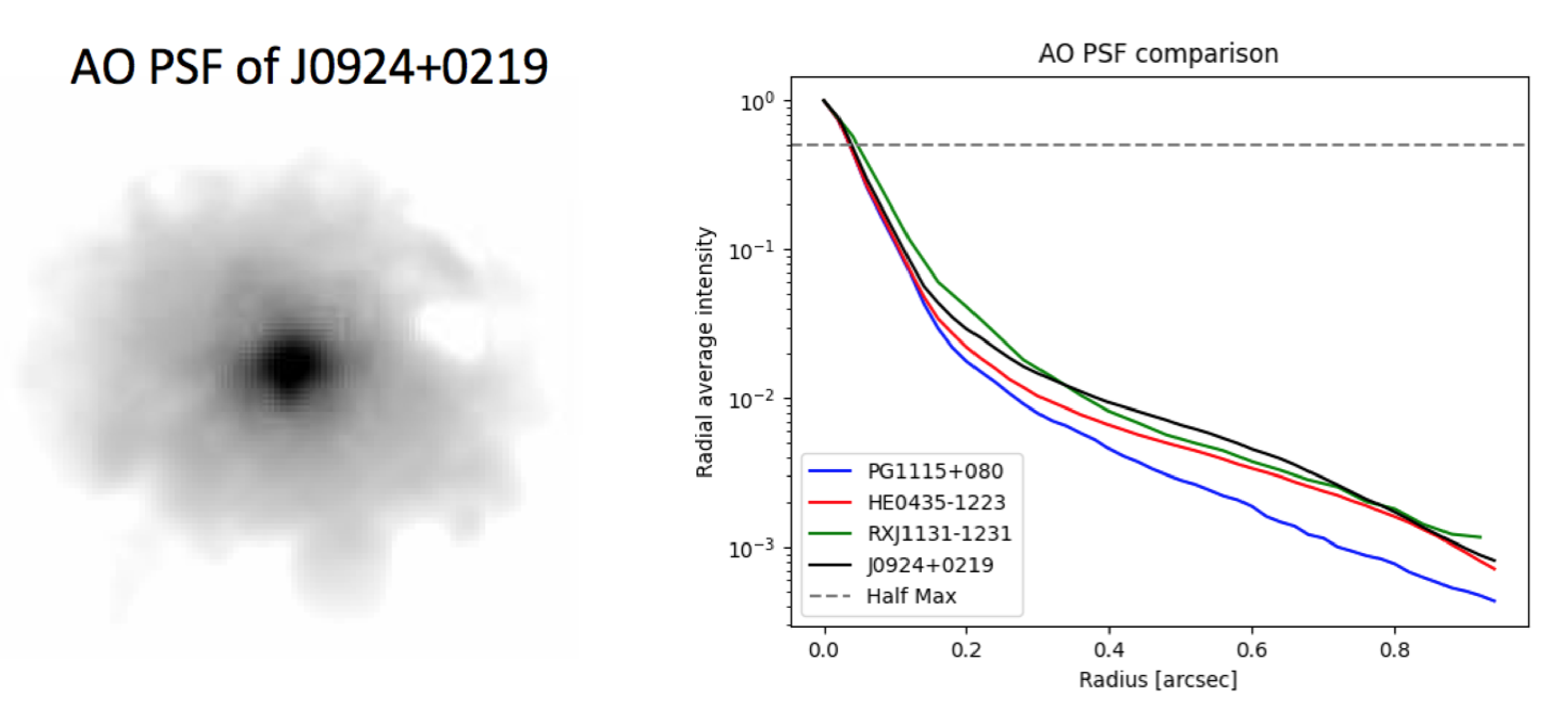}
    \caption{The left figure is the reconstructed AO PSF of \jj. The right panel is the comparison of the radial average intensity of the reconstructed PSFs from all four AO lenses from previous work \citep[][]{GChenEtal19}. All the reconstructed PSFs show core structures and extended wings.}
    \label{fig:jjAOPSF}
\end{figure*}
    
\subsection{Hubble Space Telescope Imaging}
\label{sec:HSTreduction}
We use optical and near-infrared imaging of the system obtained from the HST archive.  The archival data include NICMOS images through the F160W filter (total exposure time: 5311.52 seconds) taken with HST on November 23, 2003 and ACS/WFC images though the F814W filter (total exposure time: 2296 seconds) and F555W filter (total exposure time: 2188 seconds) taken with HST on November 18, 2003 (PID:9744, PI: C. Kochanek).
We process the data using AstroDrizzle with standard settings, which removes the geometric distortions, corrects for sky background variations, and flags cosmic-rays. The final drizzled HST images with a scale of 0.05\arcsec\ per pixel are presented in \fref{fig:aoimages}.

\subsection{Keck Adaptive Optics Imaging}
\label{sec:AOreduction}
The AO imaging was obtained at K$^\prime$-band with the Near-infrared Camera 2 (NIRC2), as part of the SHARP AO effort (Fassnacht et al., in prep). The target was observed with the narrow camera setup, which provides a roughly 10$\times$10\arcsec\ field of view and a pixel scale of 10 milliarcsec (mas). 
There are three exposures of 300 seconds on December 30, 2011, seven exposures of 300 seconds on May 16, 2012, and four exposures of 300 seconds on May 18, 2012. 
The total exposure time was 4200 seconds.
We follow our previous work \citep{GChenEtal16,GChenEtal19} and use the SHARP python-based pipeline, which performs a flat-field correction, sky subtraction, correction of the optical distortion in the images, and a coadditon of the exposures. 
For the distortion correction step, the images are resampled to produce final pixel scales of 10~mas pix$^{-1}$ for the narrow camera. The narrow camera pixels samples well the AO PSF, which has typical FWHM values of 60--90~mas.  
To improve the modeling efficiency for the narrow camera data, we perform a 2$\times$2 binning of the images produced by the pipeline to obtain images that have a 20~mas pix$^{-1}$ scale. The final HST images are presented in \fref{fig:aoimages}.

\begin{figure}
    \centering
    \includegraphics[width=\linewidth]{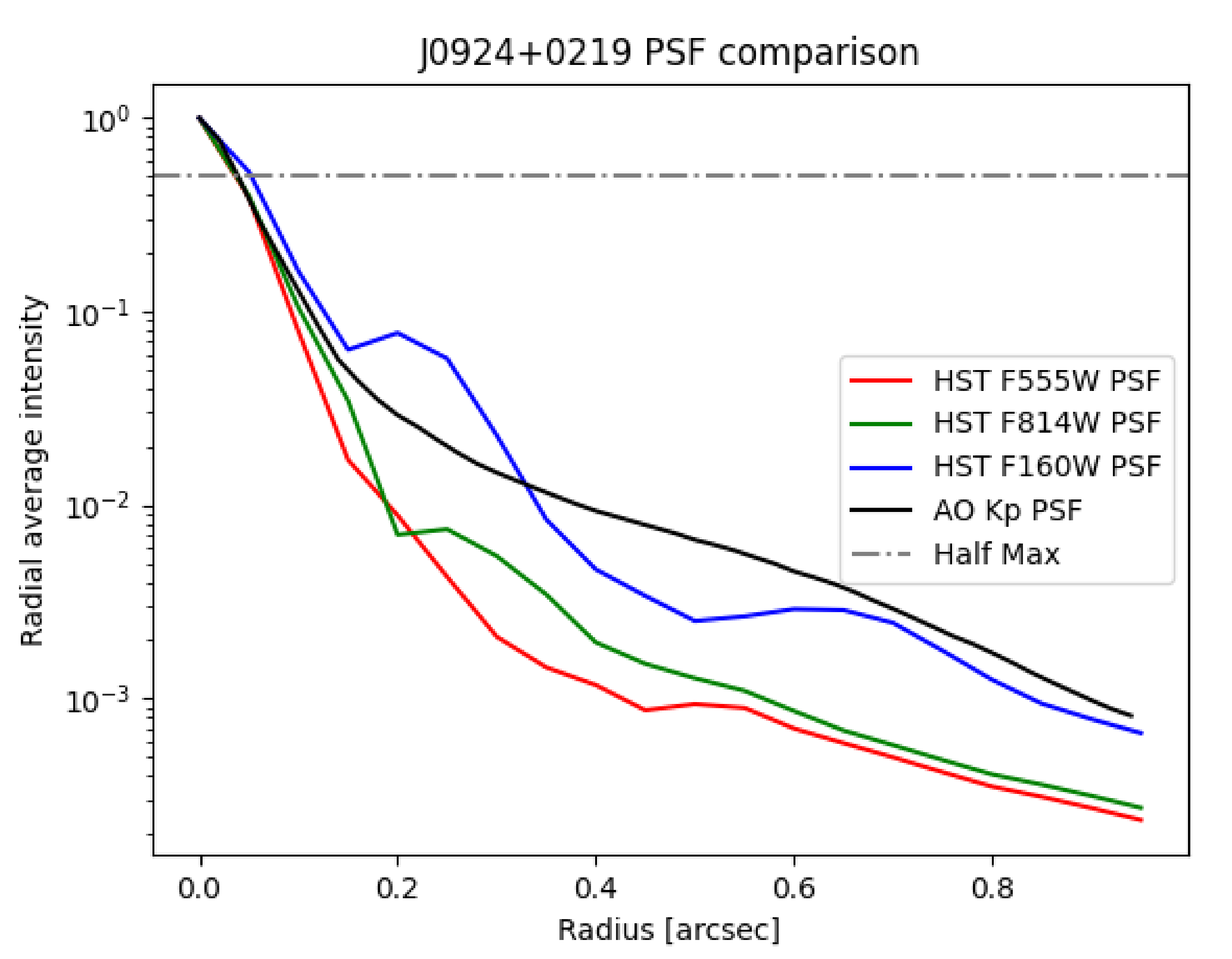}
    \caption{The comparison of the radial average intensity of the reconstructed AO PSFs and HST PSF of \jj.}
    \label{fig:jjAOHSTPSF}
\end{figure}

\section{J0924+0219 modeling}
\label{sec:jjmodeling}
We describe the PSF models in \sref{subsec:jjpsf}, lens modeling in \sref{subsec:jjlensmodeling}, kinematics modeling in \sref{subsec:jjkinematics}, and time-delay prediction model in \sref{subsec:jjtdmodel}.
\subsection{The PSF of J0924}
\label{subsec:jjpsf}
For the F160W band HST imaging, we use {\sc Tinytim} \citep{KristHook97} to generate the PSFs with different spectral index, $\hat{\eta}_{v}$, of a power-law from $-0.4$ to $-2.5$ and different focuses\footnote{The flux per unit frequency interval is $F_{\nu}=C\nu^{\hat{\eta}_{v}}$, where $\hat{\eta}_{v}$ is the power-law index and C is a constant; focus is related to the breathing of the secondary mirror, which is between $0\sim10$.} from 0 to 10. 
Given the F160W band HST imaging, we find that the best-fit to the imaging is the PSF with focus equal to 0 and spectral index equal to $-1.3$. 
We use this {\sc Tinytim} PSF as the initial guess and then apply the PSF-correction method of \citet{GChenEtal16} while modeling the F160W HST imaging. 
For the F814W and F555W bands, that were observed with the ACS with a larger field of view, we use one of the nearby bright stars as the initial guess of the PSF and apply the PSF-correction until the residuals stabilized.  
For the AO imaging, we follow the criteria described in Section 4.4.3 of \citet{GChenEtal16} and perform 9 iterative steps to create the final PSF and make sure the size of the PSF for convolution is large enough ($1.18\arcsec\times1.18\arcsec$) such that the results are stable.
The full width half maximum (FWHM) of the AO PSF is $\sim 75$ mas. We show the reconstructed AO PSF in \fref{fig:jjAOPSF} and the comparison of AO and HST PSF in \fref{fig:jjAOHSTPSF}.

\subsection{Lens imaging modeling}
\label{subsec:jjlensmodeling}
\citet{EigenbrodEtal06} first modeled this system with HST imaging and suspected that the second set of bluer arcs in F814W band (see \fref{fig:aoimages}) inside and outside the area delimited by the red arcs in F160W band could be either a second source in different redshift or a star forming region in the source galaxy. 
We examine the possibility of a second source plane existing at a lower redshift than the source ($z=1.52$) due to bluer color and find that the scenario is very unlikely, as the macro model determined by the red arc cannot reproduce a reasonable source for the blue arcs given a possible range of the source redshift from $z=0.5$ to $z<1.52$. In contrast, we do find that a star forming region can be reconstructed at the same source redshift. \citet{FaureEtal11} modeled the lens with high-resolution H and Ks imaging obtained using the ESO VLT with adaptive optics and the laser guide star system. 
They identified a luminous object, located $\sim0.3\arcsec$ to the north of the lens galaxy, but showed that it cannot be responsible for the anomalous flux ratios. 
Many studies \citep[e.g.,][]{MetcalfMadau01,BradacEtal02,DalalKochanek02,PooleyEtal12,SchechterEtal14,GlikmanEtal18,BadoleEtal20} have shown that the macro model cannot explain the flux ratio, which suggested the presence of microlensing or dark matter substructures. 
Thus, to avoid possible biases caused by flux ratios, we only use the lensed quasar positions and the extended arc to constrain the mass model, which is also the standard procedure for $H_{0}$ measurements in TDCOSMO collaboration. \citet{GilmanEtal20} also show that the presence of substructures do not bias $H_{0}$ above the percent level.
We use {\sc glee}, a strong lens modeling code to model three HST bands and one Keck AO band simultaneously \citep{SuyuHalkola10,SuyuEtal12a}.
We describe the models in the following for fitting the high resolution imaging data. 
\begin{figure*}
    \centering
    \includegraphics[width=0.8\linewidth]{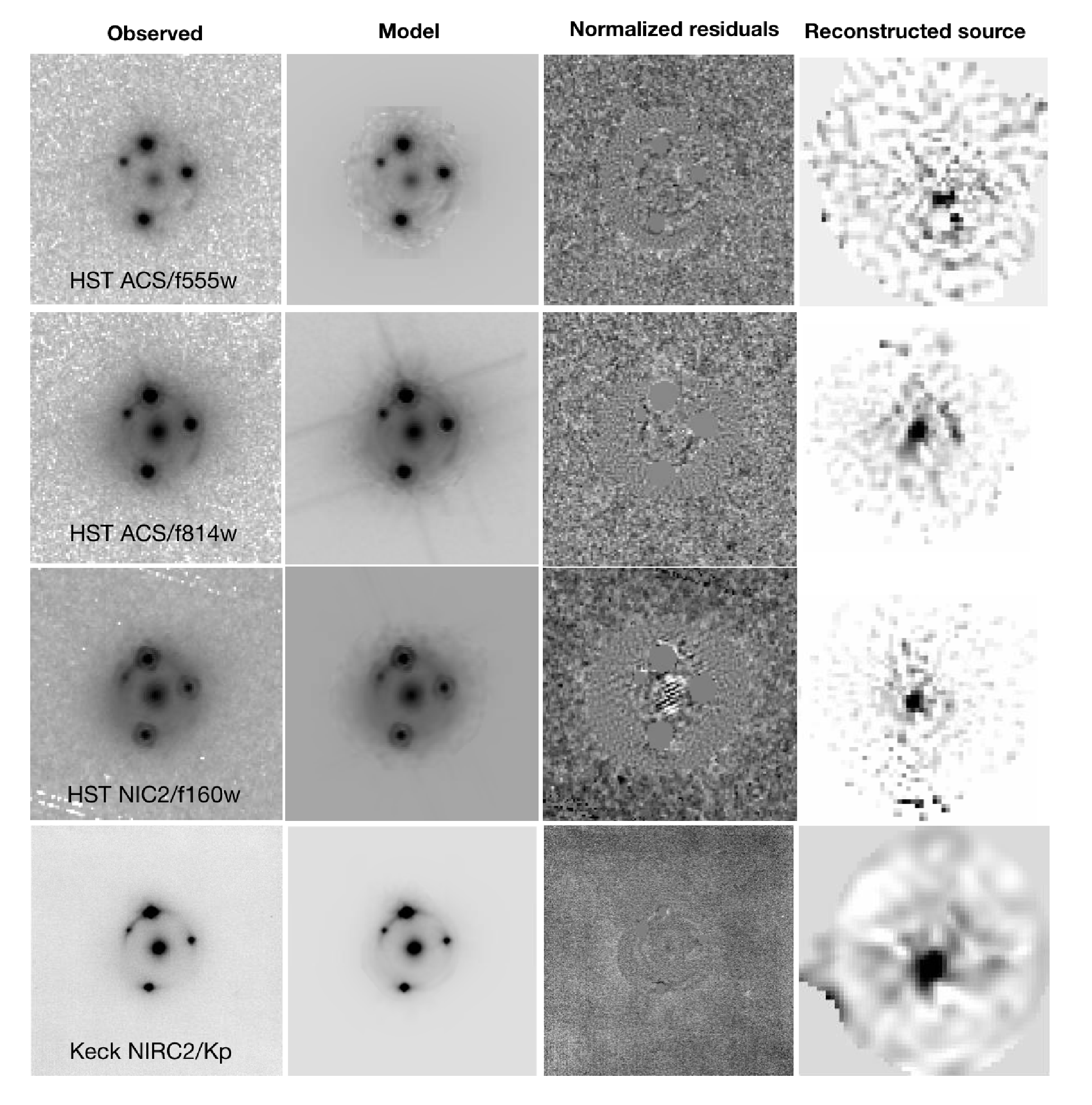}
    \caption{\jj~HST and AO image reconstruction of the most probable model with a source grid of $53\times53$ pixels. We use $59\times59$ pixels of the AO PSF and $29\times29$ pixels of the HST PSF for convolution of spatially extended images. From left column to right column: observed imaging, model imaging, normalized residuals, and reconstructed source.}
    \label{fig:J0924_residuals}
\end{figure*}
We show the imaging, models, normalized residuals, and reconstructed sources in \fref{fig:J0924_residuals}. Note that since the source in F555W band has more clumpy star forming region, the reconstructed source is less regular with small-scale structures and more noise\footnote{See also the same effect in \citet{WongEtal17}.}. In addition, the noise-overfitting problem is due to the fact that the outer region of the source plane is under-regularized, but this effect will not underestimate the uncertainty because the uncertainty will be dominated by the time-delay and velocity dispersion measurements. Besides, we model the imaging with different source resolutions and marginalize over them to control the systematics.
 
\begin{table}
\caption{Lens model parameters for power-law model} 
\label{tab:powerlawparameters1} 
\begin{center} 
\begin{tabular}{lll}  
\hline\hline 
Description & Parameter & Marginalized \\
&&Constraints\\
\hline
Lens mass distribution                & \\ 
\hline 
Centroid of G in $\theta_{1}$ (arcsec) & $\theta_{1}$\tnote{$\clubsuit$} & $3.01^{+0.08}_{-0.05}$\\ 
Centroid of G in $\theta_{2}$ (arcsec) & $\theta_{2}$ & $3.02^{+0.03}_{-0.04}$ \\ 
Axis ratio of G                       & $q_{}$ & $0.61\pm0.01$\\
Position angle of G        &$\theta$\tnote{$\spadesuit$} & $-0.04\pm0.01$\\ 
Einstein radius of G (arcsec)         & $\theta_{\rm E}$ & $0.940^{+0.004}_{-0.003}$\\
Radial slope of G                     & $\gamma$ & $2.270^{+0.007}_{-0.003}$      \\
External shear strength               & $\gamma^{\prime}$& $0.017^{+0.001}_{-0.003}$\\
External shear angle       & $\theta_{\gamma^{\prime}}$& $4.24^{+0.01}_{-0.03}$\\
\hline
\hline
\end{tabular} 
\begin{tablenotes}
{\bf Note}: The mass model parameters of power-law model. The source pixel parameters are marginalized. The confidence interval represents 1 $\sigma$ uncertainty. Position angle is counter clockwise from +x in radians.

\end{tablenotes}
\end{center}
\end{table}

\begin{figure*}
    \centering
    \includegraphics[width=\linewidth]{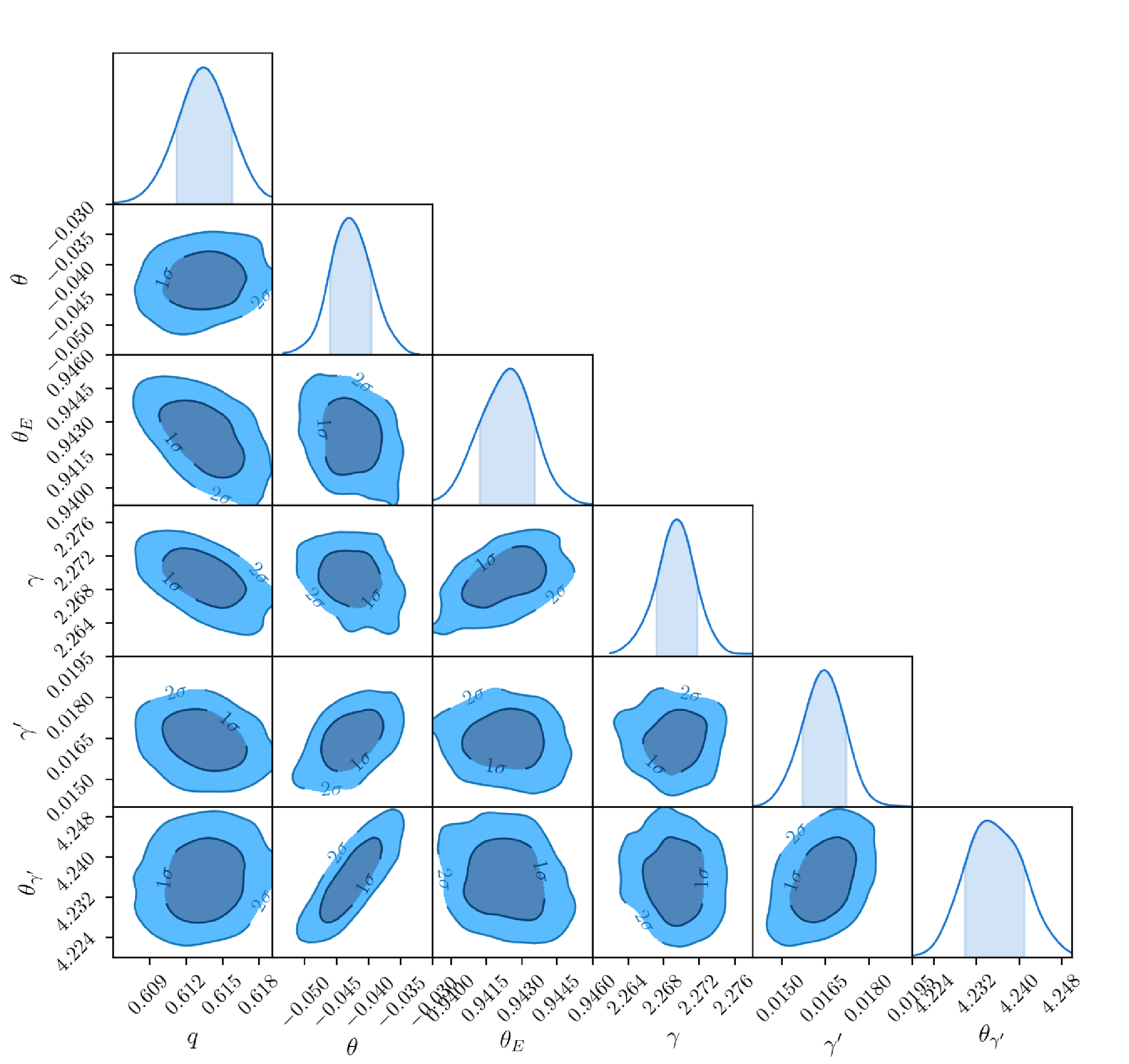}
    \caption{Marginalized mass-model parameter distributions from the \jj\ power-law lens model results. The description of the parameters are: $q$ is axis ratio of power-law mass profile, $\theta$ is the position angle of power-law mass profile, $\theta_{\rm E}$ is the Einstein radius, $\gamma$ is the slope, $\gamma^{\prime}$ is the strength of the external shear, and $\theta_{\gamma^{\prime}}$ is the orientation of the shear strength. The contours represent the $68.3\%$ and $95.4\%$ quantiles. Position angle is counter clockwise from +x in radians.}
    \label{fig:jjpowerlaw}
\end{figure*}

\begin{figure*}
    \centering
    \includegraphics[width=\linewidth]{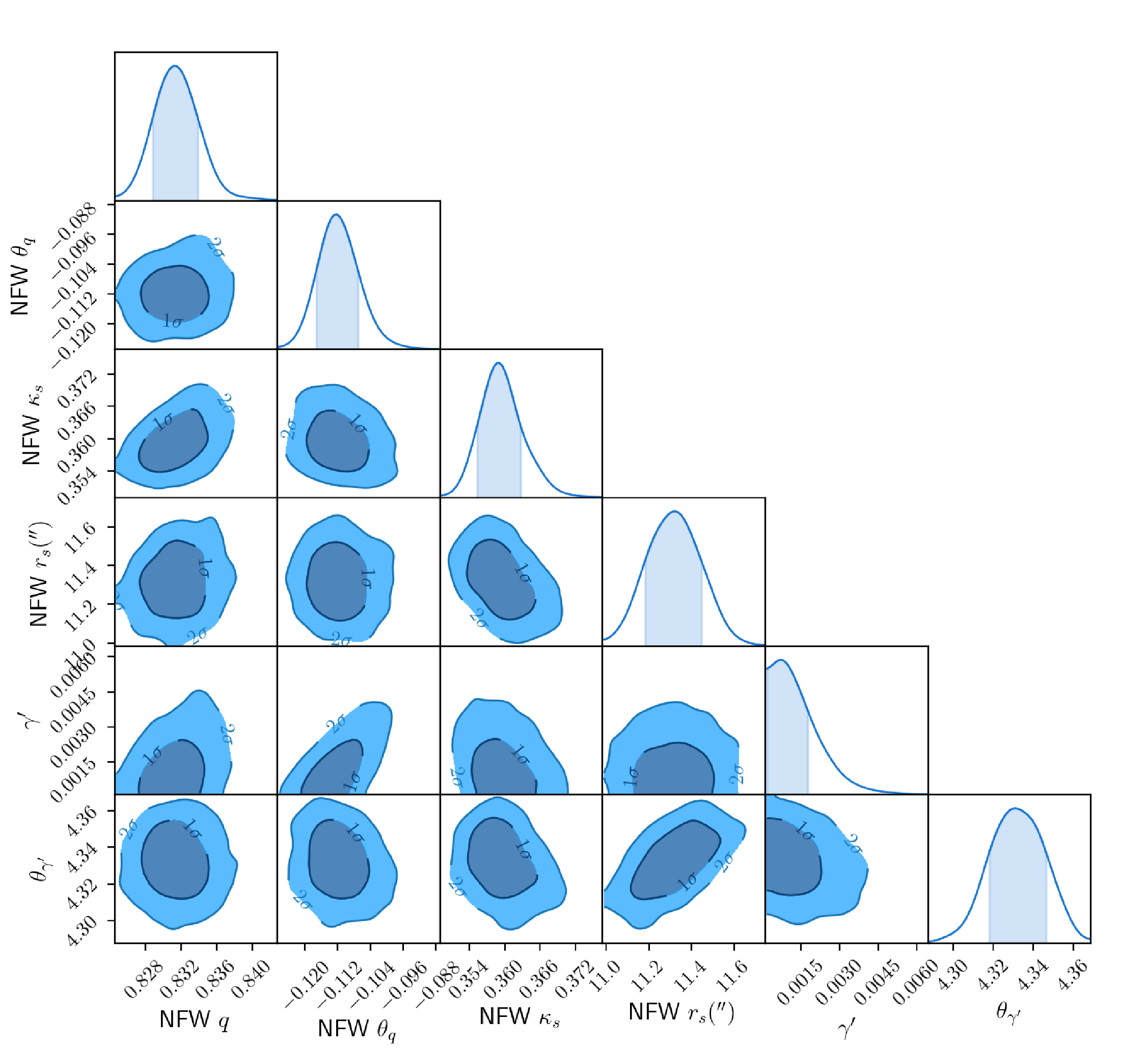}
    \caption{Marginalized parameter distributions from the \jj\ composite lens model results. NFW $q$ is axis ratio of NFW profile, NFW $\theta_{q}$ is the position angle of NFW, NFW $r_{\rm s}$ is the scale radis of NFW profile, $\gamma^{\prime}$ is the strength of the external shear, $\theta_{\gamma^{\prime}}$ is the orientation of the shear strength. The contours represent the $68.3\%$ and $95.4\%$ quantiles. Position angle is counter clockwise from +x in radians.}
    \label{fig:jjcomposite}
\end{figure*}

\begin{itemize}
    \item {\bf Power-law mass model+shear+S$\acute{\text{e}}$rsic light model}: 
    we first choose the softened power-law elliptical mass distributions \citep[SPEMD;][]{Barkana98} density profile with the softening length close to zero -- the main parameters include radial slope ($\gamma$), Einstein radius ($\theta_{\rm E}$), Position angle ($\theta_{q}$) and the axis ratio of the elliptical isodensity contour ($q$) -- to simultaneously model the extended arcs seen in the three HST bands and one AO band, and reconstruct the source structure on a pixelated grid \citep{SuyuEtal06}. The power-law model is motivated by many studies which have shown that a power-law model provides a good description of the lensing galaxies and dynamical studies for galaxy-galaxy lensing \citep[e.g.,][]{KoopmansEtal06,KoopmansEtal09,SuyuEtal09,AugerEtal10,BarnabeEtal11,SonnenfeldEtal13,CappellariEtal15,ShajibEtal21}. 
    In the modeling, we found that two concentric S$\acute{\text{e}}$rsic profiles are sufficient to describe the lensing light distribution of the HST F555W and HST F160W bands, while three concentric S$\acute{\text{e}}$rsic profiles are needed for the HST F814W band and Keck AO band. 
    Except for the parameters that describe the lens light center ($\theta_{1,\rm Light}$ and $\theta_{2,\rm Light}$), which are linked together for the light profiles, the light parameters (position angles, ellipticities, and S$\acute{\text{e}}$rsic index) are free. 
    We list all parameters in \tref{tab:powerlawparameters1} and \tref{tab:powerlawparameters2}, and show the important marginalised mass model parameters in \fref{fig:jjpowerlaw}.
    \item {\bf Composite mass model+shear+chameleon light profile}: we follow \citet{GChenEtal19} and test a composite (baryonic + dark matter) model. 
    For the dark matter component we adopt the standard NFW profile \citep{NavarroEtal96} with the following parameters:
    halo normalization (NFW $\kappa_{\rm s}$), halo scale radius (NFW $r_{\rm s}$), halo minor-to-major axis ratio (NFW $q$), and associated position angle (NFW $\theta_{q}$). This is motivated by \citet{DuttonTreu14}, who find that non-contracted NFW profiles are a good representation for the dark matter halos of massive elliptical galaxies \citep[See also][]{ShajibEtal21}.
    The baryonic component is modeled by multiplying the lens surface brightness distribution by a constant M/L ratio parameter. 
    For computational efficiency, we model the surface brightness with chameleon profile. The chameleon profile is the difference of two isothermal profiles and is a good approximation to a S$\acute{\text{e}}$rsic profile over the range of interest \citep[see details in][]{DuttonEtal11}. 
    We link the baryonic matter to the chameleon light profiles of the F160W bands because it probes the rest-frame near-infrared and thus should be the best tracer of stellar mass \citep[See also][]{GChenEtal19,WongEtal17}.
    Since the degeneracy between the wings of the AO PSF and lens light could bias the inferred baryonic component, we do not use AO lens light to infer the baryonic distribution \citep[][]{GChenEtal19}. 
    However, when combining with HST imaging, the well-known HST PSF can provide the information of baryonic distribution \citep{GChenEtal19}. Future AO imaging with AO PSF reconstructed from telemetry data can break the degeneracy and directly infer the baryonic matter without the need of HST imaging \citep[][]{GChenEtal21_AOPSF}.
    We set a Gaussian prior of $r_{s} = 15.0\pm2.0\arcsec$ based on the results of \citet{GavazziEtal07} for lenses in the SLACS sample, which encompasses the redshift of \jj. We list all parameters in \tref{tab:compositeparameters1} and \tref{tab:compositeparameters2}, and show the important marginalised parameters in \fref{fig:jjcomposite}.
\end{itemize}

\begin{table*}
\caption{Lens light model parameters for power-law model} 
\label{tab:powerlawparameters2} 
\begin{center} 
\begin{tabular}{llllll}  
\hline\hline 
Lens light as S$\acute{\text{e}}$rsic profiles\\
\hline
Description & Parameter & F555W &F814W & F160W &Keck AO\\
\hline
Centroid of S in $\theta_{1}$ (arcsec) & $\theta_{1,\rm Light}$& $3.0092\pm0.0002$& $3.0092\pm0.0002$& $3.0092\pm0.0002$& $3.0092\pm0.0002$\\
Centroid of S in $\theta_{2}$ (arcsec) & $\theta_{2,\rm Light}$& $2.9935\pm0.0002$& $2.9935\pm0.0002$& $2.9935\pm0.0002$& $2.9935\pm0.0002$\\

Axis ratio of  S1                     & $q_{\rm S1}$& $0.88^{+0.03}_{-0.04}$& $0.67^{+0.02}_{-0.03}$& $0.89^{+0.02}_{-0.01}$& $0.76\pm0.03$\\
Position angle of S1        &$\theta_{\rm S1}$ & $5.2\pm0.4$& $6.55^{+0.02}_{-0.06}$ & $3.8\pm0.1$& $-9.12^{+0.07}_{-0.08}$\\ 
Amplitude of S1                      & $I_{\rm s,S1}$\tnote{$\dagger$}&$0.2\pm0.2$&$0.071^{+0.004}_{-0.01}$&$0.669^{+0.005}_{-0.007}$&$0.41\pm0.02$ \\
Effective radius of S1 (arcsec)       & $R_{\rm eff,S1}$&$0.105\pm0.005$&$0.95\pm0.03$&$0.112\pm0.001$&$0.96\pm0.03$\\
Index of S1                           & $n_{\text{S1}}$& $0.6\pm0.1$& $0.9\pm0.1$& $1.25\pm0.02$& $0.366^{+0.005}_{-0.007}$\\

Axis ratio of  S2                     & $q_{\rm S2}$& $0.93\pm0.05$& $0.89^{+0.04}_{-0.08}$& $0.76\pm0.05$& $0.82^{+0.05}_{-0.06}$\\
Position angle of S2        &$\theta_{\rm S2}$& $0.8\pm0.1$& $6.7\pm0.2$ & $0.55\pm0.07$& $-1.7\pm0.1$\\ 
Amplitude of S2                      & $I_{\rm s, S2}$\tnote{$\dagger$}&$0.00632^{+0.00008}_{-0.0004}$&$2.1\pm0.1$&$0.023^{+0.002}_{-0.001}$&$6.3^{+1.5}_{-2.2}$  \\
Effective radius of S2 (arcsec)       & $R_{\rm eff,S2}$&$1.3\pm0.1$&$0.104^{+0.007}_{-0.004}$&$0.79^{+0.04}_{-0.07}$&$0.145\pm 0.01$\\
Index of S2                           & $n_{\text{S2}}$& $3.2^{+0.4}_{-0.6}$& $1.1^{+0.1}_{-0.1}$& $0.368\pm0.003$& $0.9\pm0.2$\\

Axis ratio of  S3                     & $q_{\rm S3}$&...& $0.52^{+0.08}_{-0.04}$&...& $0.7\pm0.1$\\
Position angle of S3        &$\theta_{\rm S3}$&... & $7.82^{+0.03}_{-0.04}$&... & $-2.9\pm0.2$\\ 
Amplitude of S3                      & $I_{\rm s, S3}$\tnote{$\dagger$}&...&$0.29\pm0.05$ &...&$0.29\pm0.05$ \\
Effective radius of S3 (arcsec)       & $R_{\rm eff,S3}$&...&$0.27\pm 0.01$&...&$0.28\pm 0.01$\\
Index of S3                           & $n_{\text{S3}}$&...& $0.6^{+0.2}_{-0.1}$&...& $0.5^{+0.2}_{-0.1}$\\
\hline
\end{tabular} 

\begin{tablenotes}[para]
{\bf Note}: The lens lights of all 4 bands share the common centroid. The source pixel parameters are marginalized and are thus not listed. The confidence interval represents 1 $\sigma$ uncertainty. Position angle is counter clockwise from +x in radians.
\end{tablenotes}
\end{center}
\end{table*}

\begin{table}
\caption{Lens mass model parameters for composite model. The Baryonic component are described by two chameleon profiles that mimic the S$\acute{\text{e}}$rsic profiles.  Each chameleon profile is composed of two cored isothermal profiles.  We label the two chameleon profiles as B1 and B2.} 
\label{tab:compositeparameters1} 
\begin{center} 
\begin{tabular}{lll}  
\hline\hline 
Description & Parameter & Marginalized \\
&&or Optimized \\
&&Constraints\\
\hline
Lens mass distribution                & \\ 
\hline 
Mass to light ratio & M/L & $12.1\pm0.2$\\ 
Centroid of B1 in $\theta_{1}$ (arcsec) & $\theta_{1,\rm B1}$\tnote{$\clubsuit$} & $3.0096\pm0.0002$\\ 
Centroid of B1 in $\theta_{2}$ (arcsec) & $\theta_{2,\rm B1}$ & $2.9906\pm0.0002$ \\ 
Axis ratio of B1                       & $q_{\rm B1}$ & $0.811^{+0.007}_{-0.008}$\\
Position angle of B1        &$\phi_{\rm B1}$\tnote{$\spadesuit$} & $-33.05\pm0.02$\\ 
Amplitude of B1                     & $I_{\rm s,B1}$\tnote{$\dagger$}&$2.72\pm0.01$ \\
core radius 1 of B1                    & $r_{\rm c,1,B1}$ & $0.105\pm0.002$\\
core radius 2 of B1                     & $r_{\rm c,2,B2}$ & $0.182\pm0.001$\\

Axis ratio of B2                       & $q_{\rm B2}$ & $0.46\pm0.01$\\
Position angle of B2        &$\phi_{\rm B2}$\tnote{$\spadesuit$} & $-34.33\pm0.02$\\ 
Amplitude of B2                     & $I_{\rm s,B2}$\tnote{$\dagger$}&$4.89\pm0.01$ \\
core radius 1 of B2                    & $r_{\rm  c,1,B2}$ & $0.020\pm0.001$\\
core radius 2 of B2                    & $r_{\rm c,2,B2}$ & $0.06\pm0.02$\\

Centroid of NFW in $\theta_{1}$ (arcsec) & NFW $\theta_{1}$\tnote{$\clubsuit$} & $2.90\pm0.03$\\ 
Centroid of NFW in $\theta_{2}$ (arcsec) & NFW $\theta_{2}$ & $3.093^{+0.005}_{-0.05}$ \\ 
Axis ratio of NFW                      & NFW $q$ & $0.83\pm0.02$\\
Position angle of NFW        &NFW $\theta_{q}$\tnote{$\spadesuit$} & $-0.11^{+0.05}_{-0.04}$\\ 
Amplitude of NFW                      & NFW $\kappa_{\rm s}$\tnote{$\dagger$}&$0.359\pm0.003$ \\
core radius of NFW                    & NFW $r_{\rm s}(\arcsec)$ & $11.3\pm0.1$\\

External shear strength               & $\gamma^{\prime}$& $0.001\pm0.001$\\
External shear angle       & $\theta_{\gamma^{\prime}}$& $4.3\pm0.1$\\
\hline
\end{tabular} 
\begin{tablenotes}[para,flushleft]
{\bf Note}: The mass model parameters of composite model. The source pixel parameters are marginalized and are thus not listed. The confidence interval represents 1 $\sigma$ uncertainty. Position angle is counter clockwise from +x in radians.
\end{tablenotes}
\end{center}
\end{table}

\begin{table*}
\caption{Lens model parameters for composite model.
} 
\label{tab:compositeparameters2} 
\begin{center} 
\begin{tabular}{llllll}  
\hline\hline 
Description & Parameter &F555W& F814W& F160W& Keck AO\\
\hline 
Lens light as S$\acute{\text{e}}$rsic profiles\\
\hline
Axis ratio of S1                     & $q_{\rm S1}$&$0.92\pm0.03$& $0.67^{+0.02}_{-0.03}$& ...& $0.75\pm0.03$\\
Position angle of S1        &$\phi_{\rm S1}$ &$4.9\pm0.2$& $6.55^{+0.02}_{-0.06}$& ...& $-9.11^{+0.07}_{-0.08}$\\ 
Amplitude of S1                      & $I_{\rm s,S1}$\tnote{$\dagger$}&$0.158\pm0.007$&$0.072^{+0.004}_{-0.01}$ &...&$0.40\pm0.02$  \\
Effective radius of S1 (arcsec)       & $R_{\rm eff,S1}$&$0.175\pm0.005$&$0.96^{+0.03}_{-0.02}$&...&$0.96^{+0.03}_{-0.02}$\\
Index of S1                           & $n_{\text{S}\acute{\text{e}}\text{rsic}, \rm S1}$&$1.69\pm0.09$& $0.86^{+0.1}_{-0.07}$& ...& $0.365^{+0.006}_{-0.007}$\\

Axis ratio of  S2                     & $q_{\rm S2}$&$0.72\pm0.04$& $0.89^{+0.05}_{-0.08}$& ...& $0.83^{+0.05}_{-0.06}$\\
Position angle of S2        &$\phi_{\rm S2}$ &$0.28\pm0.07$& $6.6\pm0.2$&...& $-1.7\pm0.1$\\ 
Amplitude of S2                      & $I_{\rm s, S2}$\tnote{$\dagger$}&$0.0046\pm0.0004$&$2.1\pm0.1$ &...&$6.3^{+1.5}_{-2.2}$\\
Effective radius of S2 (arcsec)       & $R_{\rm eff,S2}$&$2.11^{+0.08}_{-0.07}$&$0.100^{+0.007}_{-0.004}$&...&$0.15\pm 0.01$\\
Index of S2                           & $n_{\text{S}\acute{\text{e}}\text{rsic},\rm S2}$&$1.1\pm0.1$& $1.06^{+0.07}_{-0.1}$& ...& $0.9^{+0.3}_{-0.2}$\\

Axis ratio of  S3                     & $q_{\rm S3}$&...& $0.52^{+0.08}_{-0.04}$&...& $0.7\pm0.1$\\
Position angle of S3        &$\phi_{\rm S3}$ &...& $7.82^{+0.03}_{-0.04}$&...& $-2.8\pm0.2$\\ 
Amplitude of S3                      & $I_{\rm s, S3}$\tnote{$\dagger$}&...&$0.29\pm0.04$ &...&$0.28\pm0.04$ \\
Effective radius of S3 (arcsec)       & $R_{\rm eff,S3}$&...&$0.27\pm 0.01$&...&$0.27\pm0.02$\\
Index of S3                           & $n_{\text{S}\acute{\text{e}}\text{rsic},\rm S3}$&...& $0.6\pm0.2$&...& $0.6^{+0.3}_{-0.2}$\\

\hline

\end{tabular} 
\begin{tablenotes}[para,flushleft]
{\bf Note}: The lens lights of all 4 bands share the common centroid. The source pixel parameters are marginalized and are thus not listed. S1, S2 and S3 represents three different S\'{e}rsic profiles. The confidence interval represents 1 $\sigma$ uncertainty. Position angle is counter clockwise from +x in radians. The lens light parameters for the F160W band are based on chameleon profiles and are used to describe the baryonic lens mass distribution through a constant M/L ratio.  These chameleon parameter values for F160W are listed in Table 3.
\end{tablenotes}
\end{center}
\end{table*}


\subsection{Kinematic modeling}
\label{subsec:jjkinematics}

To predict the time delays under the presence of the MST, velocity dispersion information is required to constrain the normalization of the 3D de-projected mass model. 
We follow \citet{SonnenfeldEtal12} and calculate the three-dimensional radial velocity dispersion 
by numerically integrating the solutions of the spherical Jeans equation \citep[][]{BinneyTremaine87}
\begin{equation}
    \frac{1}{\rho_{\ast}}\frac{d(\rho_{\ast}\sigma_{\textrm{r}}^{2})}{dr}+2\frac{\beta_{\textrm{ani}}\sigma_{\textrm{r}}^{2}}{r}=-\frac{GM(r)}{r^{2}}, 
\end{equation}
where $M(r)$ follows either the power-law mass or composite model.
For the stellar component, we assume a Hernquist profile \citep{Hernquist90},
\begin{equation}
\rho_{\ast}=\frac{I_{0}a}{2\pi r(r+a)^3},
\end{equation}
where $I_{0}$ is the normalization term and the scale radius can be related to the effective radius by $a=0.551r_{\textrm{eff}}$.
To compare with the data, the seeing-convolved luminosity-weighted line-of-sight velocity dispersion can be expressed as
\begin{equation}
    (\sigma^{\textrm{P}}_{v})^{2}=\frac{\int_{\mathcal{A}}[I(R)\sigma_{s}^{2}\ast\mathcal{P}]d\mathcal{A}}{\int_{\mathcal{A}}[I(R)\ast\mathcal{P}]d\mathcal{A}},
\end{equation}
where $R$ is the projected radius, $I(R)$ is the light distribution, $\mathcal{P}$ is the PSF 
convolution kernel \citep{MamonEtal05}, and $\mathcal{A}$ is the aperture. 
The streaming motions (e.g. rotation) are assumed to be zero. The luminosity-weighted line-of-sight velocity dispersion is given by
\begin{equation}
    I(R)\sigma_{s}^{2}=2\int^{\infty}_{R}(1-\beta_{\textrm{ani}}\frac{R^{2}}{r^{2}})\frac{\rho_{\ast}\sigma^{2}_{\textrm{r}}rdr}{\sqrt{r^{2}-R^{2}}}.
\end{equation}

The predicted velocity dispersion can be simplified and well-approximated \citep{BirrerEtal16,BirrerEtal20,GChenEtal20} as 
\begin{equation}
\label{eq:vd_nolambda}
    (\sigma_v^{\textrm{p}})^{2}=(1-\kappaext)\lambdaint\left(\frac{\Ds}{\Dds}\right) c^{2}J(\eta_{\textrm{lens}},\eta_{\textrm{light}},\beta_{\textrm{ani}}),
\end{equation}
where 
$J$ contains the angular-dependent information
including the parameters describing the 3D deprojected mass distribution, $\eta_{\textrm{lens}}$, the surface-brightness distribution in the lensing galaxy, $\eta_{\textrm{light}}$, and the stellar orbital anisotropy distribution, $\beta_{\textrm{ani}}$. $\kappaext$ and $\lambdaint$ represents the external MST and internal MST, respectively.

We assume the anisotropy component has the form of an anisotropy radius, $r_{\textrm{ani}}$, in the Osipkov-Merritt (OM) formulation
\citep{Osipkov79,Merritt85},
\begin{equation}
    \beta_{\textrm{ani}}=\frac{r^{2}}{r^{2}_{\textrm{ani}}+r^{2}},
\end{equation}
where $r_{\textrm{ani}}=0$ is pure radial orbits and $r_{\textrm{ani}}\rightarrow\infty$ is isotropic with equal radial and tangential velocity dispersions.
In our models, we use a scaled version of the anisotropy parameter, $a_{\textrm{ani}}\equiv r_{\textrm{ani}}/r_{\textrm{eff}}$, where $r_{\textrm{eff}}=\Dd \theta_{\textrm{eff}}$, and $\theta_{\textrm{eff}}$ is the effective radius in angular units.
Note that since the LOS velocity dispersion has a degeneracy with the anisotropy parameters \citep{Dejonghe87}, we follow \citet{GChenEtal19} and marginalize the sample of $a_{\textrm{ani}}$ over a uniform distribution $[0.5,5]$.

\begin{figure*}
    \centering
    \includegraphics[width=\linewidth]{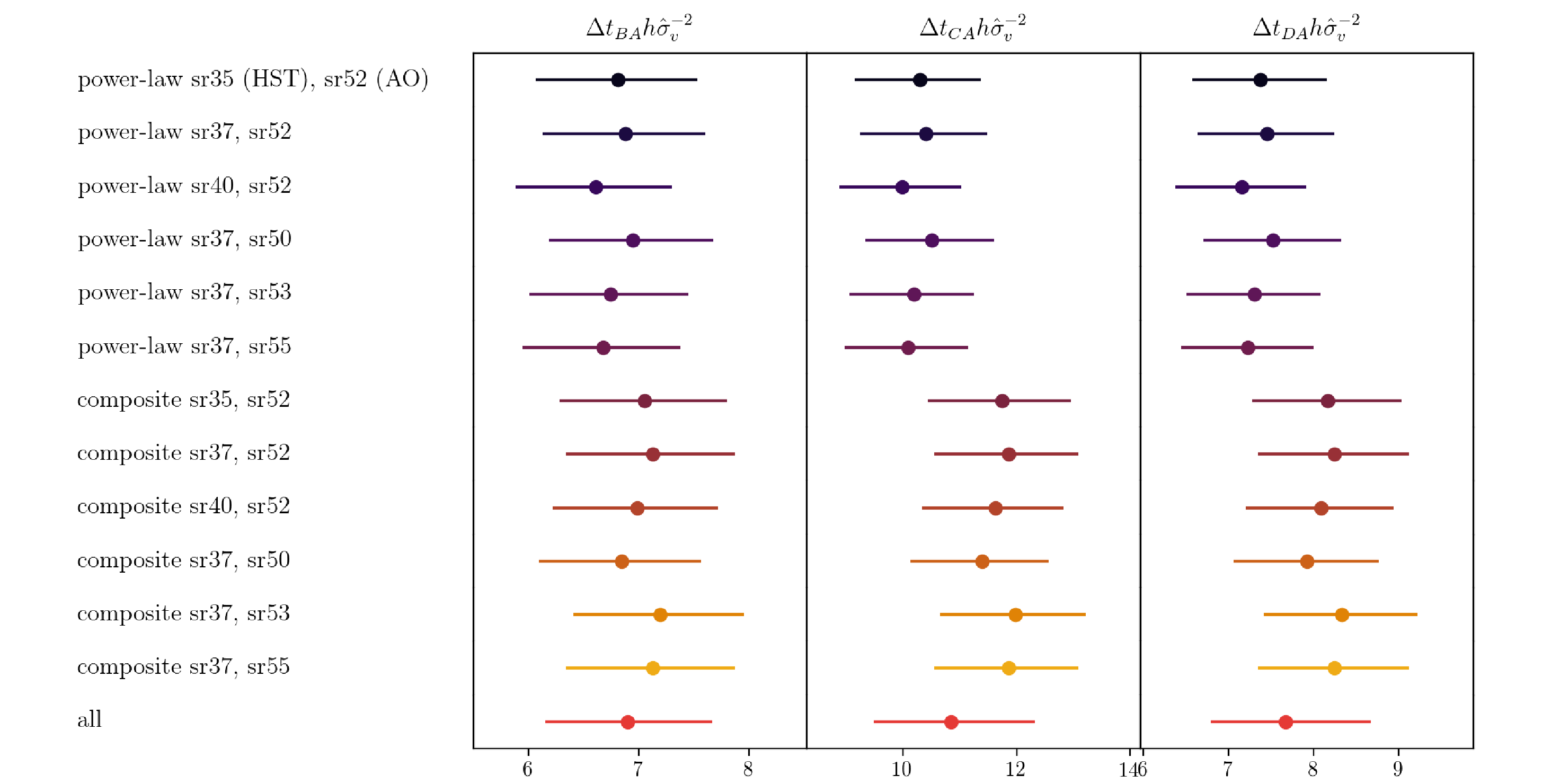}
    \caption{The predicted time delays from the power-law and composite models with different number of reconstructed source pixels for the HST imaging and AO imaging. "sr35 (HST)" represents that we use source grid with 35x35 to reconstructed the backgound source of the HST imaging. All three bands of the HST imaging share the same number of reconstructed sources.}
    \label{fig:TD_predicted}
\end{figure*}

\subsection{Time-delay prediction model}
\label{subsec:jjtdmodel}
The predicted time delay can be expressed as,
\begin{equation}
\label{eq:TD_Ddt}
\Delta t=(1-\kappaext)\lambdaint\frac{\Ddt}{c} \Delta\phi(\theta, \beta),
\end{equation}
where $c$ is the speed of light and $\theta$, $\beta$, and
$\phi(\theta)$ are the image coordinates, the source
coordinates, and the Fermat potential \citep[][]{BlandfordNarayan86} without the presence of internal or external MST respectively. 
In the case of single aperture velocity dispersion,
we can replace the MST terms ($\lambdaint$ and $\kappaext$) with \eref{eq:vd_nolambda} and the predicted time delays will directly relate to the velocity dispersion via
\begin{equation}
\label{eq:Dd}
    \Delta t=(1+z_{\rm d})\frac{\Dd }{c}\frac{\Delta\phi(\theta, \beta)}{ J(\eta_{\textrm{lens}},\eta_{\textrm{light}},a_{\textrm{ani}})}\frac{\sigma_v^2}{c^{2}}.
\end{equation} 
The MST-related terms (i.e., $\kappaext$ and $\lambdaint$) canceled out in \eref{eq:Dd}. 
Thus, the uncertainty of the predicted time delays do not depend on the uncertainty of the mass along the line of sight or transformed mass profile via MST, and only rely on the precision of the velocity dispersion measurement, the redshift of the lens, and the angular diameter distance to the lens \citep[See also similar discussion in][]{Koopmans06}. 
In other words, once the time delay and velocity dispersion are measured, the value of $\Dd$ can be determined \citep[][]{GChenEtal20}. 
When further including environmental information (which provide an estimation of $\kappaext$) and $\DsDds$ information which comes from either external datasets or assumption of a cosmological model, one can further determine $\lambdaint$ \citep[][]{BirrerEtal20,GChenEtal20} and use it to further constrain $H_{0}$ with $\Ddt$ from the population point of view \citep[][]{BirrerEtal20}. Note that \citet{BirrerEtal20} use both $\Dd$ and $\Ddt$ information to constrain $H_{0}$.

\section{Predicted time delays in $\Lambda$CDM cosmology}
\label{sec:predictedTD_LCDM}

Due to the lack of velocity dispersion measurement, we express the observed velocity dispersion as
$\sigma^{\rm ob}_{v}=\hat{\sigma}_{v}\times280~\kms$, which is created by assuming a flat $\Lambda$CDM with fixed $\Omega_{\rm m}=0.3$, $H_{0}=70~\kmsmpc$, and $\lambdaint=1$ (i.e., no internal mass-sheet transformation) in the power-law model \citep[][]{GChenEtal20}. We fold in an expected $5\%$ uncertainty of the velocity dispersion measurement and present time delay predictions under the assumption of the $\Lambda$CDM model with fixed $\Omega_{\rm m}=0.3$. 
For the velocity dispersion calculation, we assume the seeing is $1.0\arcsec$ and the aperture size is $1\arcsec\times1\arcsec$. We show the predicted time delays in \fref{fig:TD_predicted} with various source resolutions. 
When we marginalized over different source resolutions of the power-law model, the power-law model predicts
$\Delta t_{\rm BA}h\hat{\sigma}_{v}^{-2}=6.75\substack{+0.78\\-0.68}$ days, $\Delta t_{\rm CA}h\hat{\sigma}_{v}^{-2}=10.2\substack{+1.2\\-1.0}$ days, and $\Delta t_{\rm DA}h\hat{\sigma}_{v}^{-2}=7.31\substack{+0.86\\-0.74}$ days. When we marginalized over different source resolutions of the composite model, the composite model predicts
$\Delta t_{\rm BA}h\hat{\sigma}_{v}^{-2}=6.99\substack{+0.81\\-0.71}$ days, $\Delta t_{\rm CA}h\hat{\sigma}_{v}^{-2}=11.6\substack{+1.4\\-1.2}$ days, and $\Delta t_{\rm DA}h\hat{\sigma}_{v}^{-2}=8.10\substack{+0.96\\-0.82}$ days. When we marginalized power-law and composite model, we obtain
$\Delta t_{\rm BA}h\hat{\sigma}_{v}^{-2}=6.89\substack{+0.78\\-0.74}$ days, $\Delta t_{\rm CA}h\hat{\sigma}_{v}^{-2}=10.7\substack{+1.6\\-1.2}$ days, and $\Delta t_{\rm DA}h\hat{\sigma}_{v}^{-2}=7.70\substack{+0.97\\-0.91}$ days. Given the expected short time delay of this system, it will be challenging to measure the time delays within $10\%$ uncertainty.



\section{Conclusions}
\label{sec:conclusion}
In this work, we use the high resolution Keck AO imaging data, collected by the SHARP team, and deep HST WFC3 images through the F160W filter, HST ACS/WFC images though F555W filter and F814W filter to simultaneously constrain the mass distribution of \jj~lens system. When assuming a $\Lambda$CDM model with fixed $\Omega_{\rm m}=0.3$, we find that the power-law model predicts 
$\Delta t_{\rm BA}h\hat{\sigma}_{v}^{-2}=6.75\substack{+0.78\\-0.68}$ days, $\Delta t_{\rm CA}h\hat{\sigma}_{v}^{-2}=10.2\substack{+1.2\\-1.0}$ days, and $\Delta t_{\rm DA}h\hat{\sigma}_{v}^{-2}=7.31\substack{+0.86\\-0.74}$ days; the composite model (i.e., a NFW dark matter halo \citep{NavarroEtal96} plus a constant mass-to-light ratio stellar distribution) predicts
$\Delta t_{\rm BA}h\hat{\sigma}_{v}^{-2}=6.99\substack{+0.81\\-0.71}$ days, $\Delta t_{\rm CA}h\hat{\sigma}_{v}^{-2}=11.6\substack{+1.4\\-1.2}$ days, and $\Delta t_{\rm DA}h\hat{\sigma}_{v}^{-2}=8.10\substack{+0.96\\-0.82}$ days. When we marginalized over the power-law and composite model, we obtain
$\Delta t_{\rm BA}h\hat{\sigma}_{v}^{-2}=6.89\substack{+0.78\\-0.74}$ days, $\Delta t_{\rm CA}h\hat{\sigma}_{v}^{-2}=10.7\substack{+1.6\\-1.2}$ days, and $\Delta t_{\rm DA}h\hat{\sigma}_{v}^{-2}=7.70\substack{+0.97\\-0.91}$ days. Future measurements of time delays with 10\% uncertainty and velocity dispersion with 5\% uncertainty would yield a $H_0$ constraint of $\sim15$\% precision.

It is important to note that our analysis is {\sl truly} blind since the time delays and velocity dispersion are not yet measured. Once the velocity dispersion measurement and time delays are measured, the derived posteriors can be used to constrain the $H_{0}$. As part of the TDCOSMO effort, we are getting everything for this lens to have a high-quality $H_0$ measurement under the assumptions of standard NFW profile and fixed M/L ratio. These assumptions are in general supported by \citet[][]{ShajibEtal21} and are currently the standard in the TDCOSMO collaboration.
Future work with including varying mass-to-light (M/L) ratio, allowing contracted/expanded NFW profile, and adapting axisymmetric Jeans equations are worth examining the systematics when spatially-resolved kinematics data are obtained.

\section*{Acknowledgements}
GC-FC thanks Simon Birrer, Dominique Sluse, 
Aymeric Galan, and  Elizabeth Buckley-Geer for many insightful comments.
GC-FC, CDF, and TT acknowledge support by the National Science Foundation through grants NSF-AST-1907396 and NSF-AST-1906976 "Collaborative Research: Toward a 1\% measurement of the Hubble Constant with gravitational time delays",  NSF-1836016 "Astrophysics enabled by Keck All Sky Precision Adaptive Optics", and NSF-AST-1715611 "Collaborative Research: Investigating the nature of dark matter with gravitational lensing". We also acknowledge support by the Gordon and Betty Moore Foundation Grant 8548 "Cosmology via Strongly lensed quasars with KAPA". SHS thanks the Max Planck Society for support through the Max Planck Research Group, and is supported in part by the Deutsche Forschungsgemeinschaft (DFG, German Research Foundation) under Germany’s Excellence Strategy - EXC-2094 - 390783311.




\bibliographystyle{mnras}
\bibliography{AO_cosmography} 






\bsp	
\label{lastpage}
\end{document}